\documentstyle[11pt,newpasp,twoside,epsf]{article}
\def\edcomment#1{\iffalse\marginpar{\raggedright\sl#1\/}\else\relax\fi}
\reversemarginpar

\begin{document}
\title{Pulsar optical observation with the Very Large Telescope}

 \author{A. Ray}
\affil{Tata Institute of Fundamental Research, Bombay 400005, India}
 \author{P. Lundqvist, J. Sollerman}
\affil{Stockholm Observatory, SE-133 36 Saltsj\"obaden, Sweden}
 \author{B. Leibundgut}
\affil{European Southern Observatory, 
Garching bei M\"unchen, Germany}
 \author{F. Sutaria}
\affil{Inter Univ Centre for Astronomy \& Astrophysics, Pune 411007, India}

\begin{abstract}
Optical data in the V-band gathered with the
8.2m ESO Very Large Telescope (VLT) at the radio interferometric
position of PSR 1706-44 are presented. 
The pulsar is close to a bright star in projection and was not detected.
The pulsar magnitude limit must be fainter than V=24.5 for a distance
of $\leq$ 2\farcs0 from the bright star. 
In the outer gap model for an aligned rotor the optical flux should
scale with the gamma-ray flux. For pulsars which emit pulsed 
gamma-rays but are not detected in the optical bands, 
the synchrotron cutoff frequency for the tertiary photons
must be well below the optical frequencies
and the magnetic and spin axes may be misaligned.
\end{abstract}

\section{Optical observation of PSR1706-44 \& pulsar radiation models}

Optical detection of pulsars
aid in the development
and constraining of theoretical models of pulsar electromagnetic radiation.
PSR1706-44 belongs to the set of seven $\gamma$-ray pulsars detected by
EGRET (Thompson et al. 1996). 
In the optical it has not been detected yet. 
Detection of the optical counterparts of radio and gamma-ray pulsars
is often complicated by field crowding by other stars in the optical bands.
The field of this pulsar was observed with the 
{\it VLT}-UT1 on
August 19, 1998 (SV phase). 
The details of observation, analysis and theoretical
implications are given in Lundqvist et al (1999).
Chakrabarty and Kaspi (1998) (CK98)
using the radio position of the pulsar 
as summarised in their paper 
estimate that the pulsar should lie $2\farcs7$ away from the star 1.
The combined error in position of the optical counterpart of
the radio pulsar from various sources is $1\farcs0$.
Measuring the background
at a distance of $\sim 2\farcs7$ from nearby bright
Star 1 in the VLT data shows that an artificial star with $V=25.5$ 
can be detected at more than 3$\sigma$ level. 
To estimate how bright a star one could hide in the PSF
of Star 1 we subtracted artificial stars from this position until a
hole appeared in the background. It is possible to hide a
point source ($V=25.0$) at a distance of $\leq 2\arcsec$ from the Star 1.
As an upper limit for a pulsar this close to Star 1 we claim $V=24.5$.
\begin{figure}
\plottwo{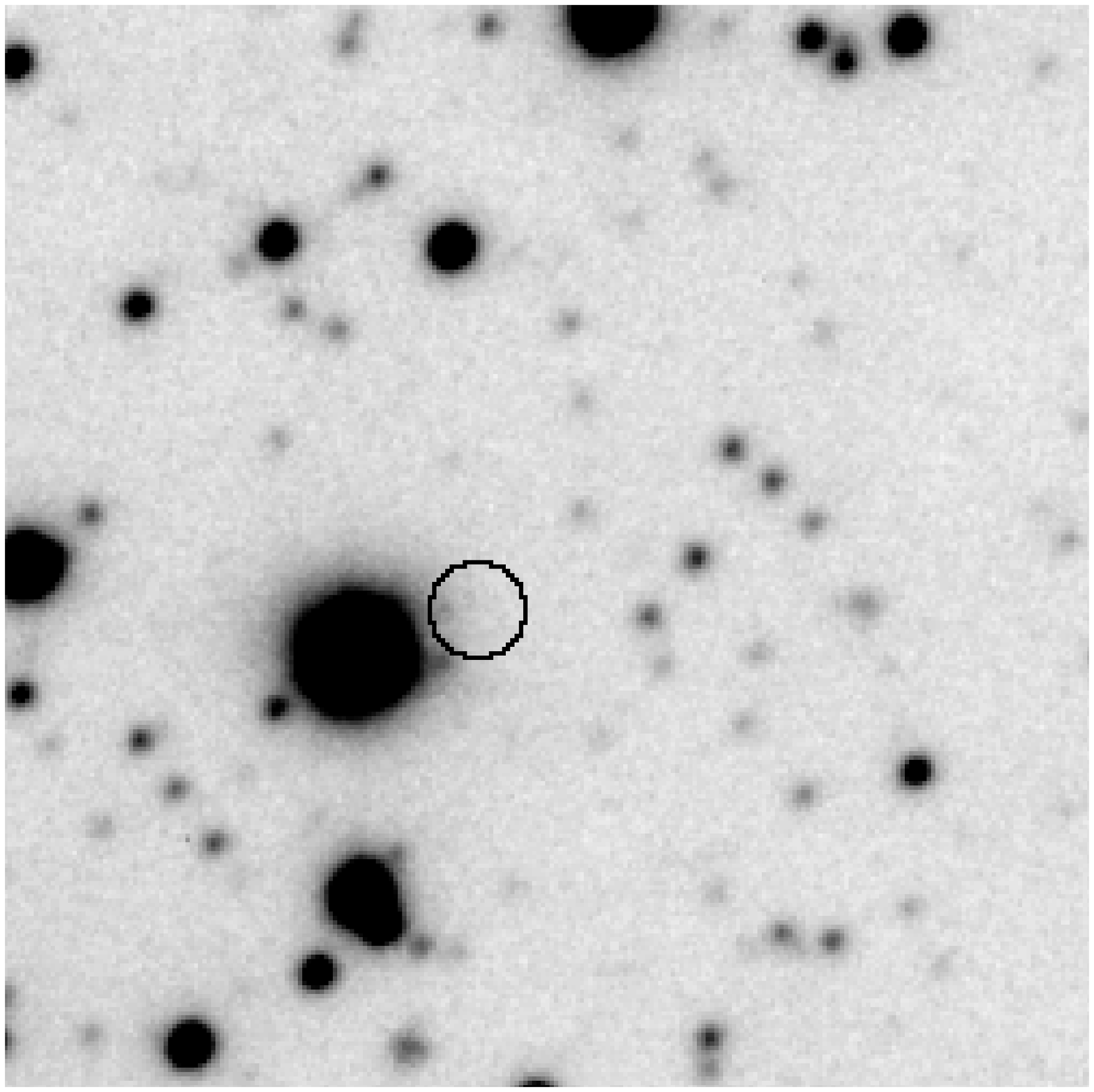}{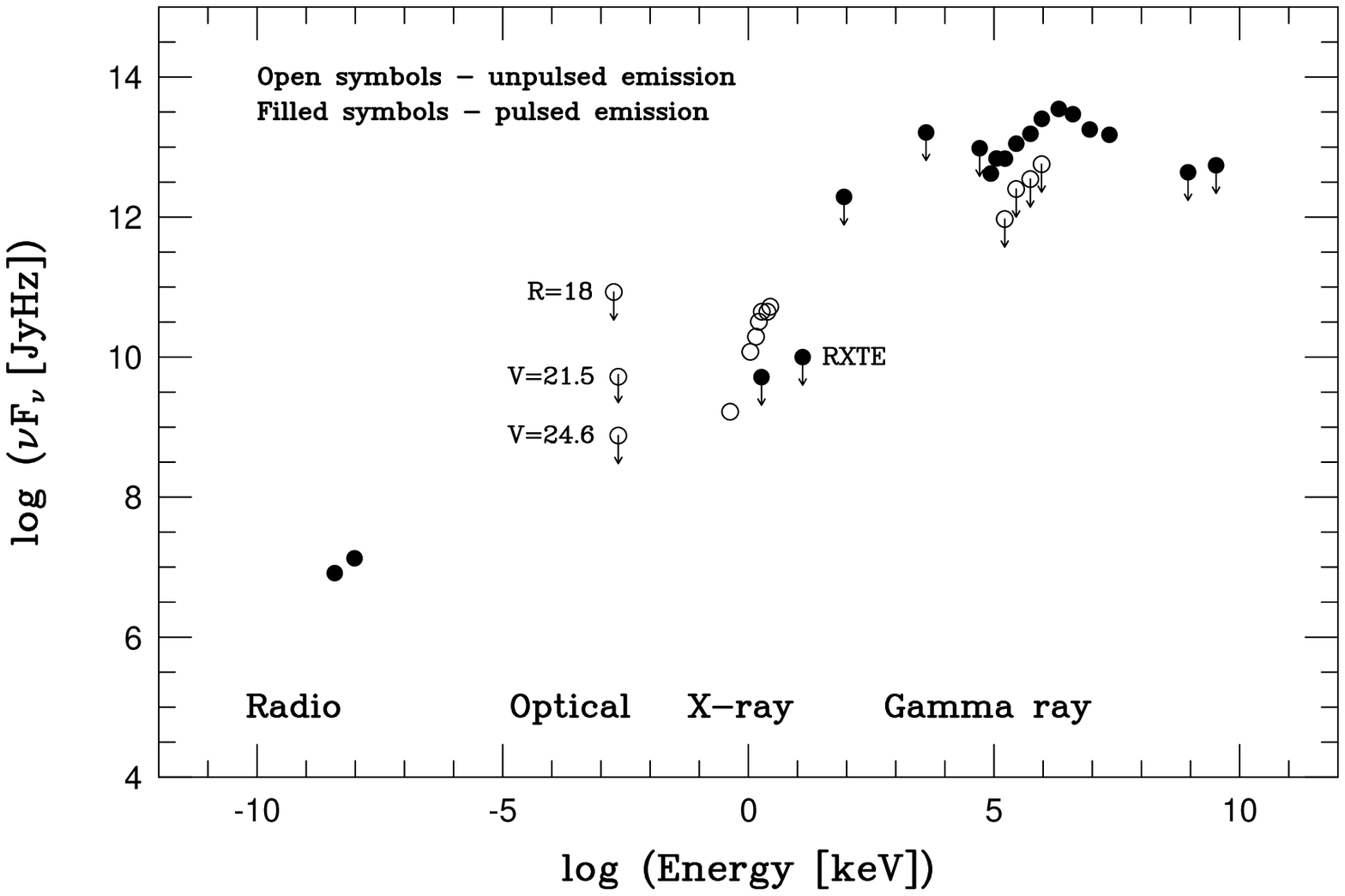}
\caption{
{\bf Left:} The expected position of PSR1706-44 
inside the 1\farcs0 error circle
centered 2\farcs7 away from Star 1.
{\bf Right:} Multiwavelength spectrum 
from data compilations of Thompson et al. (1996, 1999). 
The points marked `RXTE'
and `R' are from Ray et al (1999) and CK98 respectively. 
The VLT observation results marked V should
bracket the dereddened upper limits:
$V=24.6$ (minimum $A_{\rm V} = 0.9$ and
the pulsar outside the PSF of Star 1) and  $V=21.5$ assumes
$A_{\rm V} = 3$.
}
\end{figure}
%
%
The optical fluxes may be correlated with the
gamma-ray photon fluxes in outer gap
models (see e.g. Usov (1994) and Cheng, Ho \&
Ruderman (1986a, 1986b)) for Vela-like pulsars. 
Assuming $F_{\nu}$ to be the same in both the $V$ and $R$ bands, the magnitudes
predicted by the outer gap models are $R \la 20.0$ and $V \la 19.8$. Our faint
$V$ limit, in comparison to the predictions of the standard outer-gap
model, scaled from gamma-ray flux ($V \la 19.8$), requires a low frequency
cutoff in its synchrotron emission spectrum.
If the magnetic axis inclination with respect to spin axis
$\chi \ga \pi/4$, the synchrotron cutoff frequency is $\sim 10^{13}$ Hz
and in that case the flux of optical radiation may be very small.

\end{document}